\documentstyle[aps,prl,multicol,epsf]{revtex}
\epsfclipon
\tightenlines
\begin{document}
\draft
\title{
Rare-event induced binding transition of heteropolymers
}
\author{Lei-Han Tang$^{(a)}$ and Hugues Chat\'e$^{(b)}$}
\address{
$^{(a)}$Department of Physics, Hong Kong Baptist University, 
Kowloon Tong, Kowloon, Hong Kong\\
$^{(b)}$CEA --- Service de Physique de l'Etat Condens\'e, 
Centre d'Etudes de Saclay, 91191 Gif-sur-Yvette, France
}
\date{\today}
\maketitle
\begin{abstract}
Sequence heterogeneity broadens the binding transition 
of a polymer onto a linear or planar substrate. This effect is 
analyzed in a real-space renormalization group scheme designed to 
capture the statistics of rare events.
In the strongly disordered regime, binding initiates
at an exponentially rare set of ``good contacts''. 
Renormalization of the contact potential yields
a Kosterlitz-Thouless type transition in any dimension. 
This and other predictions are confirmed by extensive
numerical simulations of a directed polymer interacting with a
columnar defect.
\end{abstract}
\pacs{64.60.Ak,64.70.Pf,68.35.Rh,87.15.Aa}

\begin{multicols}{2}

The binding transition of a polymer onto another extended object,
such as a second polymer, a membrane, or an interface, is of interest
in a variety of physical circumstances\cite{fish84}.
The problem has been studied in the context of wetting in two 
dimensions\cite{forg86,dhv},
the depinning of a flux line from a columnar defect in 
type-II superconductors\cite{nv93},
the denaturation of double-stranded DNA molecules in solution
\cite{ps70,hwa}, and the localization of a copolymer at a
two-fluid interface\cite{ghlo}.
In surface physics, formation of facets on vicinal surfaces
can be viewed as a bound state problem involving surface steps
\cite{pv98}.

The mathematical framework for the homopolymer binding transition
is well-established. The binding transition of a heteropolymer 
(i.e., a disordered sequence of two or more letters) is expected
to be broadened by fluctuations in the monomer contact energy with the
target object.
But a quantitative characterization of the broadening effect, particularly
when the disorder fluctuations are strong, is still largely missing.
At present, the only settled case is that of
a directed heteropolymer interacting with a columnar 
defect, where weak variations in the contact energy have been shown to
have no effect on the transition, 
provided the transverse dimension $d$ is between one and three
\cite{forg86,dhv}. 
For $d=1$, Forgacs {\it et al.}, using a field-theoretic method,
found that weak disorder only introduces subleading order corrections
to the singular part of the free energy and 
does not smear out the specific heat jump at the transition\cite{forg86}. 
This conclusion was put into doubt by Derrida {\it et al.}, who performed
a perturbative renormalization group (RG) analysis
and numerical calculations\cite{dhv}. The problem has since been 
re-examined with conflicting results\cite{hwa,lassig}.

In this Letter we report a novel RG approach to the 
heteropolymer binding, focusing on the statistics of
rare attractive segments that initiate the transition.
The frequency for these ``good contacts'', in the case of a random
sequence, is described by an exponential tail in
the distribution of the contact (free) energy whose decay
rate $q$ depends on temperature. 
When $q$ reaches a critical value, the energy gain at these segments
is sufficient to offset the logarithmic entropy cost to form a
bound phase. This offers an extremely robust and generic
mechanism of heteropolymer binding transition at strong disorder
in any dimension.
The critical behavior around the transition is analyzed in a real-space
RG calculation. Combined with extensive simulation results
on the directed random heteropolymer, we conclude that, in the
disorder dominated regime, the binding transition becomes
infinite order, bearing many of the characteristics of
the Kosterlitz-Thouless-Berezinskii (KTB) transition in 
the two-dimensional XY model\cite{kost73}.

To set the scene, let us recall the directed polymer (DP) problem on
a two-dimensional square lattice $(r,t)$ with a columnar defect at 
$r=0$. For simplicity, we restrict
the transverse displacement of the DP at each step $\Delta t=1$ 
to three possible values, $\Delta r=-1,0,1$. 
A step on the defect at $r=0$
picks up an energy $\eta(t)$ which depends on the location of the
contact (or monomer index) $t$. In the case of a random heteropolymer,
the contact energies $\eta(t)$ are taken to be
identically distributed and independent from each other.

The above model is easily generalized to the case of
a directed walk on a $(d+1)$-dimensional hypercubic lattice,
interacting with a line defect.
The role of extra transverse dimensions is best appreciated
in the ``necklace'' representation\cite{fish84}, where
the partition sum of a DP of length $L$ is expressed as,
\begin{equation}
Z(L)=\sum_{0\leq t_1<t_2<\ldots<t_n\leq L}\exp[-H(\{t_i\})/T].
\label{Z(L)}
\end{equation}
Here $t_i$ is the monomer index of the $i$th contact, and 
\begin{equation}
H(\{t_i\})=\sum_i\eta(t_i)+\sum_i u(t_{i+1}-t_i).
\label{necklace-H}
\end{equation}
The ``pair potential'' $u(t)$ is obtained by 
summing over thermal paths in the bulk connecting
two successive contacts separated by a distance $t$.
It has the generic form,
\begin{equation}
u(t)/T=f_0 t +\beta(d)\ln t +\mu+O(t^{-1}),
\label{pair-potential}
\end{equation}
where $f_0$ is the reduced free energy per unit length in the bulk,
$\beta(d)=1+{1\over 2}|d-2|$
is a universal exponent characterizing the
first return probability of a random walker in $d$ dimensions,
and $\mu$ is a model-specific parameter\cite{note1}.

As emphasized by Fisher\cite{fish84}, 
the contact point representation (\ref{Z(L)})
unifies a large class of polymer binding problems
including, under certain approximations,
polymers that are self-avoiding or interacting with a surface.
The linear term in $t$ in (\ref{pair-potential})
drops out when the ratio $Z(L)/Z_0(L)$ is considered,
where $Z_0(L)=\exp(-f_0L)$ is the partition function of a free polymer.
The dominant interaction at large distances is given by
the logarithmic term. Since the density of contacts vanishes at a continuous
transition, the critical properties are governed by the value of $\beta$, 
which serves to define universality classes. 
For this class of models, it suffices to consider the DP problem
at $d\leq 2$, which is the case we focus on below.

Previous analytical studies, based on a perturbative treatment of
the contact potential, ran into difficulties when disorder becomes 
relevant\cite{forg86,dhv,lassig}.
To elucidate how strong disorder modifies the transition,
we examine first an extreme case
where the contact potential is infinite everywhere except on
a small number of randomly distributed sites of density $A$.
The reduced contact potential $x_i=\eta(t_i)/T$ on these sites is assumed to
be negative and exponentially distributed with 
a probability density function (PDF) $p(x)=q\exp(qx)$, where
$q$ is the decay rate. 
For $q<1$, the one-contact partition sum 
$z=\sum_{1\leq i\leq L}\exp(-x_i)$ of a polymer of length
$L\gg 1/A$ is dominated by the strongest binding site (i.e., glassy).
The typical value of the potential on this site is 
$x_m\simeq -q^{-1}\ln(LA)$. 
The excess reduced free energy of a
single contact is then estimated to be, 
\begin{equation}
\Delta F/T\simeq x_m+\beta\ln L=(\beta-q^{-1})\ln L+const.
\label{one-contact}
\end{equation}
Hence, for $q<q_c=1/\beta$, a sufficiently long polymer
will make use of the rare occurence of relatively large
binding energy (of order $\ln L$) on a single monomer to form the bound phase.
By setting $\Delta F\simeq T$, we estimate a ``localization
length'' $\xi_\parallel\simeq \exp(const/|q-q_c|)$.
Polymers of length $L<\xi_\parallel$ are typically free, while those
with $L>\xi_\parallel$ are typically bound.

Although real polymers seldom (if at all) qualify for the 
above description at the monomer scale, it turns out that the
exponential tail can be self-generated under coarse-graining.
This should come across as no surprise by noting the following.
(i) For a random heteropolymer,
a sequence of consecutive negative $\eta$'s 
appears spontaneously. The binding energy is proportional
to the sequence length $n$ which follows the exponential distribution.
(ii) Above the transition temperature $T_{\rm c}$, 
$\langle Z^q\rangle/Z_0^q$ diverge with
polymer length for $q>q(T)$, while those with $q<q(T)$ remain finite.
This implies that the PDF of $Z$ 
has a power-law tail $Z^{-q(T)-1}$ towards large values.
Equivalently, the PDF of the reduced free energy,
$F/T=-\ln Z$, decays exponentially at rate $q(T)$.

The task now is to devise a suitable framework to track the
renormalization of the contact energy distribution due to multiple
contacts. We have succeeded in achieving this end under
a real-space RG scheme. The main ideas are outlined below.
The contact point representation (\ref{necklace-H}) defines
a one-dimensional lattice gas model. 
Consider a segment of $b$ sites along the chain
which is mapped to a single site under a block transformation.
The site is occupied (i.e., contains a contact)
if at least one of the $b$ sites on the 
segment is occupied, and empty otherwise.
In the former case, the Boltzmann weight of a contact,
$w_i=\exp[-\eta(t_i)/T]$, transforms as,
\begin{equation}
w=\prod_{i=1}^b (1+w_i)-1.
\end{equation}
Rescaling $t\rightarrow bt$ reduces $w$ by a factor
$b^{\beta}$ due to the logarithmic interaction.  
Hence the following iterative RG equation
is seen to hold, 
\begin{equation}
1+\tilde w={1\over b^\beta}\prod_{i=1}^b (1+w_i)+1-{1\over b^\beta}.
\label{iter_w}
\end{equation}

For $b=2$, Eq. (\ref{iter_w}) coincides with the iterative equation
for the partition function $Z=1+w$ on the Berker lattice given by 
Derrida {\it et al.}\cite{dhv}
A full treatment of the problem can be formulated in terms of the
PDF $P(x)$ of the reduced contact free energy $x\equiv -\ln(1+w)$.
From the distribution of the sum $y=\sum_i x_i$,
\begin{mathletters}
\begin{equation}
Q(y)=\int dx_1\cdots dx_b\delta\Bigl(y-\sum_{i=1}^b x_i\Bigr)
\prod_{i=1}^bP(x_i),
\label{def-Q}
\end{equation}
we obtain the distribution of $\tilde x\equiv -\ln(1+\tilde w)$,
\begin{equation}
\tilde P(\tilde x)=Q[\psi(\tilde x)]{d\psi\over d\tilde x}.
\label{P-tilde}
\end{equation}
\label{functional-iter}
\end{mathletters}
Here $\psi(\tilde x)=\tilde x-\beta\ln b
-\ln\bigl[1+(b^{-\beta}-1)\exp(\tilde x)\bigr]$
is obtained from (\ref{iter_w}) by solving for $y$.

In the limit $b\rightarrow 1$,
Eqs. (\ref{functional-iter}) define a RG flow in the space of distributions.
Since $Q(x)$ is a convolution of $P(x)$, it is convenient to introduce the 
Laplace transform,
$$\hat P(s)=\int_{-\infty}^0 dx e^{-sx}P(x).$$
Collecting terms to the first order in
$dl\equiv\ln b$, we obtain,
\begin{equation}
{d\hat P(s)\over dl}=\hat P(s)\ln\hat P(s)-\beta s[\hat P(s)-\hat P(s-1)].
\label{P_flow}
\end{equation}

Consider first the homopolymer problem described by 
$\hat P(s)=\exp(-as)$. The parameter $a$, which defines an effective
free energy on a given scale, satisfies,
\begin{equation}
da/dl=a+\beta\bigl(1-e^a\bigr).
\label{flow-a}
\end{equation}
The stable fixed point at $a=0$ corresponds to the unbound phase
where the contact potential renormalizes to infinity.
Transition to the bound phase occurs at $a=a_{\rm c}<0$.

The PDF $P(x)=Aq\exp(qx)+(1-A)\delta(x)$ considered in the single-contact
model above corresponds to
\begin{equation}
\hat P(s)=\cases{
1+As/(q-s),&$s<q;$\cr
\infty,&$s\geq q.$}
\label{P-eigen}
\end{equation}
Setting $A=0$, we obtain a family of fixed-point
functions of (\ref{P_flow}), parametrized by $q$. 
Writing $\hat P(s)=\langle Z^s\rangle$, where
$Z=1+w$ represents a certain restricted partition function,
we see that these fixed-points describe correctly the
behavior of the system above the binding transition.

It can be shown that, within the class of functions that
have a power-law singularity at $s=q$, (\ref{P-eigen}) is the only
one that is renormalizable\cite{note2}.
we have derived a two-parameter RG flow equation from
(\ref{P_flow}) by matching terms that diverge as $s\rightarrow q$.
For $A\ll 1$, the result reads,
\begin{mathletters}
\begin{eqnarray}
dq/dl&=& -{1\over 2}qA,\\
dA/dl&=& (1-q\beta)A+{2-q\over 2q}A^2.
\end{eqnarray}
\label{2-parameter}
\end{mathletters}

\begin{figure}
\epsfxsize=8cm
\epsfbox{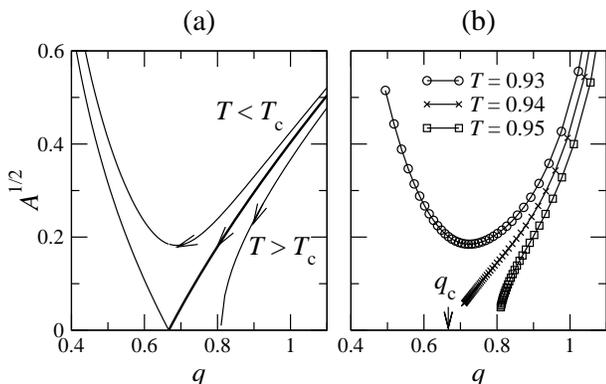}
\caption{RG flow from Eqs.~(\ref{2-parameter}) (a) and from full iteration
of the distribution on the Berker lattice (b) (see text).
}
\label{fig1}
\end{figure}

The RG flow obtained by integrating
(\ref{2-parameter}) at $\beta=3/2$ is illustrated in Fig.~\ref{fig1}(a),
where $q$ is plotted against $A^{1/2}$ in analogy with
the KTB transition.
The critical manifold ($T=T_{\rm c}$) is indicated by the thick line.
For $T>T_{\rm c}$, the flow ends on the $A=0$ axis
at a $q>q_{\rm c}=1/\beta$. In contrast, below but close to $T_{\rm c}$,
$A$ first decreases, reaches a minimum value and then increases again.
Integrating the flow equation near $q=q_{\rm c}$
and $A=0$ yields an exponentially diverging correlation length
$\xi_\parallel\simeq \xi_0\exp(\alpha|T-T_{\rm c}|^{-1/2})$. 
As in the KTB transition, the free energy per unit length has
an essential singularity when approaching $T_{\rm c}$ from below,
\begin{equation}
f_s\simeq T/\xi_\parallel =(T/\xi_0)\exp\bigl(-\alpha|T-T_{\rm c}|^{-1/2}\bigr).
\label{f-sing}
\end{equation}

We have checked the above predictions against (i) full iteration of
Eqs. (\ref{functional-iter}) at $b=2$ and $\beta=3/2$
(Berker lattice with a branching ratio $n=2\sqrt{2}$),
and (ii) transfer matrix calculation in (1+1) dimensions.
In both cases, the contact energy is taken to be $\eta=u_0+2g^2+2g\epsilon$,
where $u_0$ is the strength of the contact potential at the 
homopolymer binding transition,
and $\epsilon$ is a random variable satisfying the normal distribution.
With this choice, the transition temperature of the
annealed problem (as defined by $F_{\rm a}\equiv -T\ln\langle Z\rangle$)
is fixed at $T_{\rm a}=1$. The strength of the disorder can be tuned by 
varying $g$.

To probe the effective contact potential on a given scale $L$ on the
square lattice, we have computed the partition sum $Z(L)$
over all directed paths that start at $(0,0)$ and end at any point
on the $t=L$ line.
The quantity $x(L)\equiv -\ln[Z(2L)/Z(L)]$, computed for a given
realization of the disorder, yields a single-scale contact 
potential\cite{note2}. 
Figure 2(a) shows the integrated distribution $\Pi(x)$
for two temperatures at $g=2$ and a sample size $10^5$.
The exponential tail quickly develops as $L$ increases. 
At $T=T_{\rm a}$ (top), the tail decays as $\exp(x)$.
The pre-exponential amplitude decreases as a power of the 
length. At $T=T_{\rm c}$ (bottom), the decay rate of the tail
approaches the critical value $q_c=2/3$,
while the amplitude changes much slower as compared to
higher temperatures. In both cases, the typical value of $x(L)$
approaches ${1\over 2}\ln 2$ (dotted lines) expected
for a uniform repulsive potential.
For comparison, the integrated distribution
of $x=-\ln Z$ on the Berker lattice, obtained through numerical iteration
of (\ref{functional-iter}), are shown in Fig. 2(b). A similar trend is
observed, though the data spans over a much bigger range and is much
less noisy.

\begin{figure}
\epsfxsize=8.2cm
\epsffile{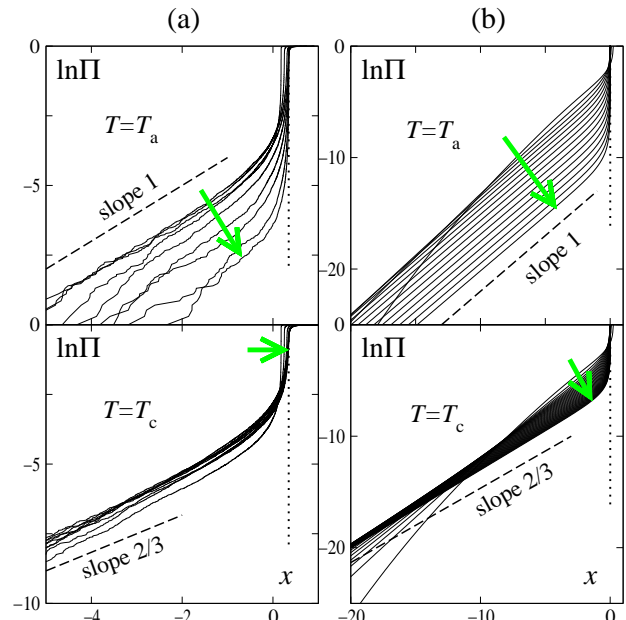}
\caption{Evolution of the integrated PDF $\Pi(x)$ as $L$ increases 
in powers of 4 (in the direction of the thick arrows).
(a) Square lattice with periodic boundary conditions and a lateral size of
$2\times 1024$ sites ($g=2$, $10^5$ samples).
(a) Berker lattice at $n=2\sqrt{2}$ and $g=1$.
}
\label{fig2}
\end{figure}

Using the Berker lattice data, we have fitted the tail of
$\Pi(x)$ to the exponential form $A\exp(qx)$.
The evolution of $A(L)$ and $q(L)$ in the critical region
is shown in Fig.~\ref{fig1}(b).
Each data set corresponds to an increasing series of $L$
in powers of four (from right to left) at a given temperature. 
The resemblance to the RG flow diagram in Fig.~\ref{fig1}(a) is apparent. 
Therefore the two-parameter flow equations (\ref{2-parameter}),
although only approximate in construct, correctly captures the
asymptotic renormalization of the tail.

\begin{figure}
\epsfxsize=\linewidth
\epsfbox{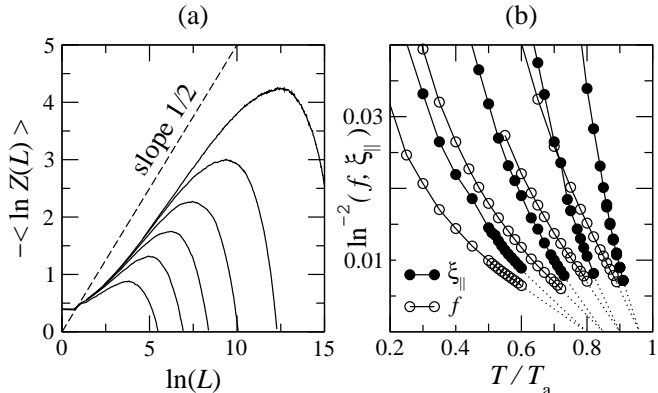}
\caption{Transfer matrix results for the square lattice.
(a) Mean free energy $-\langle \ln Z\rangle$ 
as a function of $L$ for $g=3$ and 
$T=0.15, 0.25, 0.35, 0.45, 0.55$ and 0.65 (from bottom to top).
(b) Variation of $\xi_\parallel$ and $f$ with $T$ for $g=1.5, 2, 2.5$ and 3
(from right to left).
}
\label{fig3}
\end{figure}

On the square lattice, the localization length $\xi_\parallel$ 
in the bound phase can be determined from the mean free energy
$-\langle\ln Z(L)\rangle$ which goes through a maximum
at $L=\xi_\parallel$, as shown in Fig. 3(a).
For a polymer of length $L<\xi_\parallel$, $-\langle \ln Z\rangle$ 
increases with $L$ and approaches the asymptotic behavior
${1\over 2}\ln L$ (dashed line) of a purely repulsive line.
Thus, even at $T<T_{\rm c}$, a sufficiently short polymer (or a 
short segment of a long polymer) is typically repelled by
the target object. The attraction, due to rare events, is felt
(in a typical realization of the disorder) only when $L>\xi_\parallel$. 
The asymptotic behavior at $L\gg \xi_\parallel$ is given by
$\langle \ln Z(L)\rangle\simeq -fL$, where $f(T)$ is the reduced 
free energy per unit length in the thermodynamic limit.
The variation of $\xi_\parallel$ and $f$ with $T$, as shown in 
Fig.~\ref{fig3}(b), is in agreement with Eq.~(\ref{f-sing}).
Extrapolating the curves to the horizontal axis (as indicated by
the dashed lines), we obtain $T_{\rm c}/T_{\rm a}=0.96$, 0.9, 0.85, and
0.8 for $g=1.5$, 2, 2.5, and 3, respectively. 
These values agree with the analysis based on the evolution of the 
tails of the distribution. Details of our study will be published elsewhere.

In summary, we propose that the random heteropolymer binding transition
at strong disorder is induced by rare fluctuations of the
contact potential. A real-space RG scheme is presented to analyze the
evolution of the PDF of the contact potential under successive block 
transformations. Within a two-parameter Ansatz for the tail of
the distribution, a RG flow similar to that of the KTB transition
is derived. The analysis yields an exponentially diverging
correlation length and an essential singularity for the free energy
as $T_{\rm c}$ is approached from the bound side. These predictions,
which hold in all dimensions when disorder is relevant, are
confirmed by extensive numerical calculations on the
Berker lattice and on the square lattice in (1+1) dimensions.

Finally, we comment briefly on the relationship between our work and
previous numerical studies\cite{dhv,hwa,lassig} of the DP binding 
transition on the square 
lattice, which were done on binary disorder with $\eta=\rho\pm\Delta$. 
By looking at the evolution of $P(x)$ at $T=T_{\rm a}$, we discover
that, even for $\Delta=\infty$ (while keeping $\rho-\Delta$ finite),
the tail develops at a much slower rate than for gaussian disorder. 
Since $d=1$ is marginal with regard to the relevance 
of weak disorder, we expect results reported previously to be
influenced, to a lesser or greater extent, 
by crossover effects. Indeed, the transition
temperatures $T_{\rm c}$ determined in these studies, using methods different
from ours, are either indistinguishable or only slightly below 
$T_{\rm a}$ of the corresponding annealed problem. We leave a 
detailed discussion of this and other issues to a 
future publication.

We thank the CEA Saclay (LHT) and the 
Hong Kong Baptist University (HC) for hospitality, where
the work was performed. The research is supported in part
by the Hong Kong/France PROCORE program, and by the Hong Kong
Baptist University (FRG/99-00/II-53).

\end{multicols}


\begin{references}

\bibitem{fish84} M. E. Fisher, J. Stat. Phys. {\bf 34}, 667 (1984).

\bibitem{forg86} G. Forgacs, J. M. Luck, Th. M. Nieuwenhuizen,
and H. Orland, Phys. Rev. Lett. {\bf 57}, 2184 (1986).

\bibitem{dhv} B. Derrida, V. Hakim, and J. Vannimenus, J. Stat. Phys.
{\bf 66}, 1189 (1992).

\bibitem{nv93} D. R. Nelson and V. M. Vinokur, Phys. Rev. B {\bf 48},
13060 (1993).

\bibitem{ps70} D. Poland and H. A. Scheraga, 
{\it Theory of Helix-Coil Transitions in Biopolymers},
(Academic Press, NY, 1970).

\bibitem{hwa} D. Cule and T. Hwa, Phys. Rev. Lett. {\bf 79}, 2375 (1997).

\bibitem{ghlo} T. Garel, D. A. Huse, L. Leibler and H. Orland,
Europhys. Lett. {\bf 8}, 9 (1989).

\bibitem{pv98} A. Pimpinelli and J. Villain,
{\it Physics of Crystal Growth}, (Cambridge University Press, 1998).

\bibitem{lassig} H.~Kallabis and M.~Lassig, Phys.~Rev.~Lett. {\bf 75}, 
1578 (1995).

\bibitem{kost73} J. M. Kosterlitz and D. J. Thouless, J. Phys. C {\bf 6},
1181 (1973); J. M. Kosterlitz, J. Phys. C {\bf 7}, 1046 (1974);
V. L. Berezinskii, Zh. Eksp. Teor. Fiz. {\bf 61},
1144 (1971) [Sov. Phys. JETP {\bf 34}, 610 (1972)].

\bibitem{note1} The $\mu$ term in (\ref{pair-potential}) can be absorbed
in $\eta$. It influences the transition temperature but not
the critical properties.

\bibitem{note2} L.-H. Tang and H. Chat\'e, to be published.

\end{references}
\end{document}